%% file: main.tex
\documentclass[submission,copyright,creativecommons]{eptcs}
\providecommand{\event}{TAV-WEB 2010} 
\usepackage{breakurl}             
\usepackage[utf8]{inputenc}
\usepackage[T1]{fontenc}
\usepackage{url}
\usepackage{graphicx}
\usepackage{listings}

\title{Browser-based Analysis of Web Framework Applications}

\author{Benjamin Kersten and Michael Goedicke
\institute{paluno -- The Ruhr Institute for Software Technology\\
University of Duisburg-Essen\\ 
Essen, Germany}
\email{\{benjamin.kersten, michael.goedicke\}@s3.uni-due.de}
}

\def\titlerunning{Browser-based Analysis of Web Framework Applications}
\def\authorrunning{B. Kersten \& M. Goedicke}

\begin{document}

\maketitle

\begin{abstract}
\input{abstract}

\end{abstract}

\section{Introduction}
\label{sec:introduction}

\input{introduction}

\section{Scenario}
\label{sec:scenario}

\input{scenario}

\section{Approach}
\label{sec:approach}

\input{approach}

\section{Case Study}
\label{sec:casestudy}

\input{casestudy}

\section{Discussion}
\label{sec:discussion}
\input{discussion}

\section{Related Work}
\label{sec:related}

\input{relatedwork}

\section{Conclusion and Future Work}
\label{sec:conclusion}
\input{conclusion}

\bibliographystyle{EPTCS/eptcs}
\bibliography{IEEEtran/IEEEabrv,bib/tavweb}{}

\end{document}

%% file: abstract.tex
Although web applications evolved to mature solutions providing sophisticated
user experience, they also became complex for the same reason. Complexity
primarily affects the server-side generation of dynamic pages as they are
aggregated from multiple sources and as there are lots of possible processing
paths depending on parameters. Browser-based tests are an adequate instrument to detect
errors within generated web pages considering the server-side process and path
complexity a black box. However, these tests do not detect the cause of an error
which has to be located manually instead. This paper proposes to generate metadata 
on the paths and parts involved during server-side processing to facilitate
backtracking origins of detected errors at development time. While there are
several possible points of interest to observe for backtracking, this paper focuses user
interface components of web frameworks.

%% file: introduction.tex
Sophisticated web applications do not consist of static web pages any more, but
usually make use of advanced functionality such as dynamic user
interaction or partial page updates. The benefit of this evolution is the
producibility of mature web applications with a wide range of possible features
and desktop-like user interaction. However, one drawback is the complexity
with respect to testing and locating of errors.  

\begin{figure}[t]
	\centering
		\includegraphics[width=0.9\textwidth]{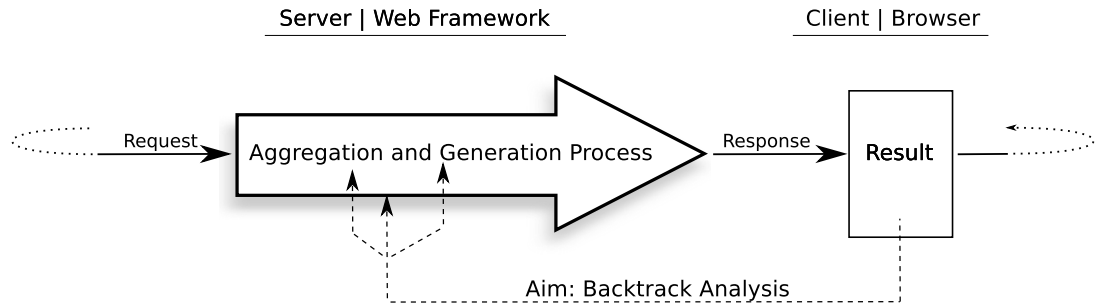}
		\caption{Modern web frameworks encompass a complex aggregation and generation
		process to produce results. This paper presents an approach to backtrack
		involved server-side parts when an element is selected within the browser. }
	\label{fig:generationProcess} 
\end{figure}

Figure \ref{fig:generationProcess} gives an overview of the coarse application
flow when utilizing web frameworks: 
Incoming requests are processed by the web framework to generate dynamic web
pages that are sent to the client in response. 
This generation process can be considered complex since request types,
parameters, application states, and session states can trigger different paths
of processing. 
Thus, the result displayed within the browser might be different than
expected. 
Furthermore, generated pages are typically aggregated from multiple involved
sources complicating the mapping of generated artifacts to their origin.  
While the details of the process complexity and affected parts are
described more detailed in section \ref{sec:scenario}, it is for now 
important to notice that it influences the tests and analyses of web applications. 

To test such web applications, there are different 
types of tests for different layers, all appropriate to their specific purpose. 
One meaningful approach is to perform tests  
within the browser, considering the generated web page as a final overall result
displayed to the user. 
The server-side process generating this page is considered
a black box and output results of input parameters are just
compared to reference values. 
Considering the complexity of the generation process mentioned above, the main
benefit of this approach is that the whole process is tested with its special
cases. 
For instance, if a requested page is expected to be processed by unit A within
the process, but is actually processed by unit B due to a parameter C triggering
a different path of processing, then an error could occur.
Therefore, unit tests may be hard to compose and cover combinations of states
and parameters. 

However, even if the necessity of browser based tests is considered, these
tests do only detect errors, but do not detect the cause of an error.  
When an error occurs, a developer usually has to locate it by guessing the
responsible part from his knowledge on the application and framework. 
That might be a time consuming effort for non-standard processings, e.g. if a
certain parameter or application state triggered a special case as described
above. 

The paper aims to provide analysis support by backtracking affected sources
from elements within the browser as depicted in figure
\ref{fig:generationProcess}. 
More precisely, selecting an element within the generated page should allow to
obtain information on its source and affected parts during server-side
processing. 
The challenge of this idea and reason why this is not possible yet is the 
one-way direction of processing as indicated by the process arrow. 
It encompasses several steps each processing data from different sources to
generate results for subsequent steps, but information on affected sources and
processing units is lost. 
Finally, the response sent as result to the client
does not contain any information on affected sources and parts of that generation process. 

The presented approach generates metadata on points of interest and transmits
it to the client to be used as basis for backtracking. 
Generally, points of interest can be any parts during server-side processing the
developer is interested in to locate errors. 
For instance, if an error occured with data displayed in the web page, points
of interest would be any data specific parts such as model updates, attached
data sources or data queries. 
This paper focuses on analyzing user interface (UI) widgets. 
Nevertheless, the approach is applicable to different points of interest. 

It is important to notice that this approach does not aim to replace exising
test types and techniques, but to enhance them by providing improved analysis. 
For instance, existing browser based tools could be used to detect errors,
while this approach can be used afterwards to locate the origins of these
errors.

The paper is structured as follows: 
Section \ref{sec:scenario} refines the problem and generation process
complexity while referring to the chosen scenario of UI widgets. 
Section \ref{sec:approach} presents the approach on how to backtrack source
information and affected server parts when selecting elements within the
browser. 
In section \ref{sec:casestudy}, the approach is applied to the scenario,
showing the feasibility with an implementation and working sample. 
Alternatives and problems of the approach are briefly discussed in section
\ref{sec:discussion}. 
Section \ref{sec:related} presents related work before the paper is concluded
in section \ref{sec:conclusion}.

%% file: scenario.tex
Section \ref{sec:introduction} announced web framework UI components to be the
point of interest for the scenario of this paper. 
More precisely, the paper deals with the Java Server Faces
(JSF) \cite{JSF12Spec, SchalkBurns06JSFCompleteReference12} web framework as
example, being part of the Java Enterprise Edition (JEE) \cite{JEE5Spec}
specification. As part of JEE, JSF is also a specification, for which different  
implementations and several extensions exist. 
This paper makes use of the reference implementation and the RichFaces
framework \cite{Katz08Richfaces}, where the latter facilitates advanced
Asynchronous Javascript and XML (AJAX) \cite{Zakas07ProfessionalAjax,
Powell08AjaxCompleteReference} communication as well as sophisticated UI widgets as extension.

JSF with RichFaces is a representative web framework solution as it
provides typical features and tasks of a web framework. 
Amongst others, this encompasses a huge set of reusable UI widgets and
processing of standard web tasks such as conversion, validation or state
management. 
\begin{figure}[t]
	\centering
		\includegraphics[width=1.0\textwidth]{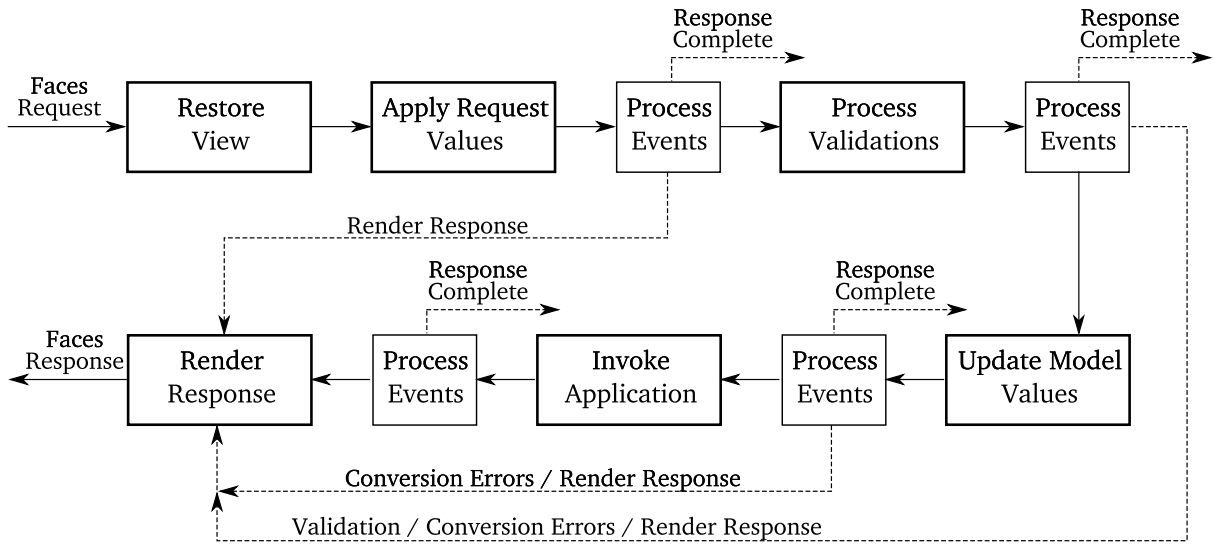}
		\caption{The lifecycle of the JSF web framework to process requests. }
	\label{fig:jsfLifecycle}
\end{figure}
Figure \ref{fig:jsfLifecycle} shows the lifecycle of JSF, handling these
standard tasks. 
Refering to the UI component processing within the JSF lifecycle, the first
stage restores UI components of preceding requests if available and the last
stage renders UI components. More details and intermediate stages are deferred
to section \ref{sec:casestudy}.

\begin{figure}[t]
	\centering
		\includegraphics[width=1.0\textwidth]{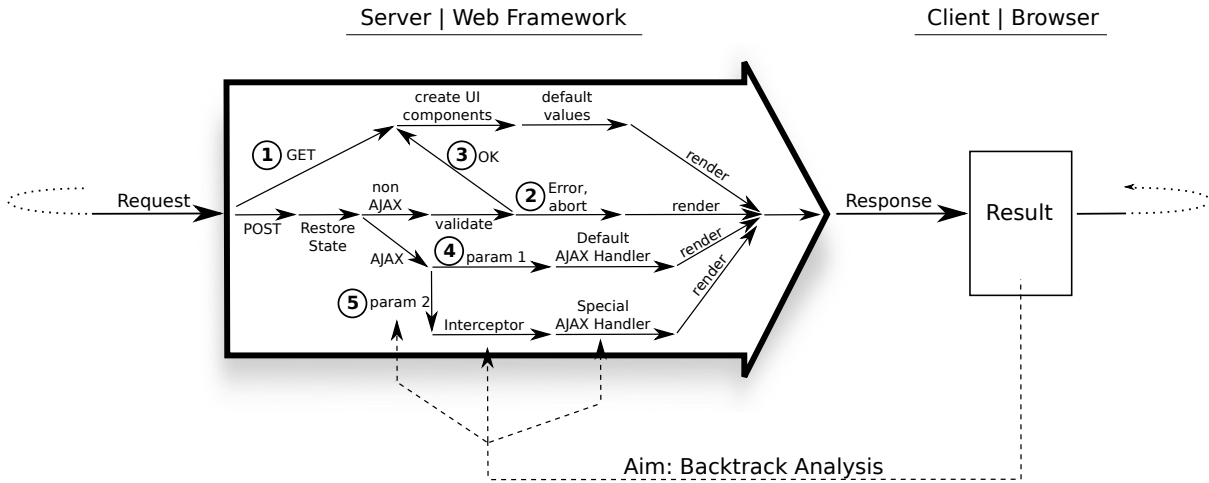}
		\caption{The JSF lifecycle can be forked at several points resulting in lots
		of possible paths for a request. Selecting an element
		within the browser should allow to track the processed path for efficient
		issue analyses.}
	\label{fig:generationProcessPaths}
\end{figure}

Although this process seems to be simple and straightforward, it can be forked
at several points causing lots of different paths within this process. 
Refining figure \ref{fig:generationProcess} which drafted the complex
generation process as black box, figure \ref{fig:generationProcessPaths} illustrates a selection of
possible paths within the process. 
For instance, some subsequent requests could be processed on different paths as
follows: 
An initial request is sent as HTTP GET \cite{HTTP_RFC}, does not contain a
session id, creates a simple view of UI components according to the requested page, applies default
values and renders the components as final HTML result. 
This also includes aggregation of data from multiple sources, e.g. the JSF page
defining the components and an external data source for their current values.  
A second request posts a formular, which fails validation due to missing
values and causes a corresponding response. 
The formular is corrected and sent again, passes validation and navigates to
another page as depicted by step 3, entailing a similar path as in the first
request. 
In a fourth step, an AJAX request is sent to perform a partial page update. 
Parameters are processed by a DefaultAjaxHandler and Renderer to generate a
response. 

Except for skipping some steps, all these requests were processed according to
the JSF lifecycle process depicted in figure \ref{fig:jsfLifecycle}. 
However, any request took a different path within this process due to
certain parameters. 

Now, a developer currently developing a part of the web application could
assume that a page is processed according to path four, i.e. he expects the
DefaultAjaxHandler to be triggered. 
However, a certain application state or parameter such as \emph{param 2} could
be intercepted and trigger a SpecialAjaxHandler instead. 
Thus, the result displayed within the browser could be different than expected,
e.g. because the SpecialAjaxHandler added additional data and styles to a
widget. 
To locate the cause of this behavior, the developer could debug the complete
process to reveal the intercepted parameter as the cause of this issue. 

This paper suggests a different approach to assist analysis: 
When the defective element is selected within the browser, the developer shall
obtain information on this element, such as passed parameters and involved 
handlers or renderers. 
In this sample, \emph{param 2}, the Interceptor and SpecialAjaxHandler would
be displayed for the request whereas the DefaultAjaxHandler does not appear. 
That helps developers to understand the generation process of selected elements
for faster analysis and locating of errors.  
Furthermore, affected lines of code can be displayed and highlighted 
within the Integrated Development Environment (IDE) to prevent
manual search within source files.

%% file: approach.tex
Sections \ref{sec:introduction} and \ref{sec:scenario} presented the aim to
generate metadata during server-side processing of web requests for being able
to backtrack issues to their sources. 
This induces several questions: 
(1) What is ment by metadata and what does it encompass? 
(2) How is that metadata collected, generated and transfered to the client? 
And finally (3) how is the metadata used on client-side to backtrack server
side sources?

The first question depends on the point of interest the developer wants to
observe. 
For instance, metadata for analysis of displayed business data could contain
used data sources and queries whereas metadata on application states could
contain pre and post values of application variables. 
In general, common data such as called methods and classes, affected lines
of code and parameter values could be useful. 
Refering to the chosen sample of JSF UI components, metadata
additionaly encompasses how components are created, how values are applied,
converted and validated as well as the rendering of components. 
In any case, required metadata has to be identified once to include it in the
next step. 

The key concept of this approach, i.e. the generation of required metadata
during server-side processing of requests is done with aspect oriented
programming (AOP) \cite{Laddad03AspectJinAction}.
AOP allows to define pointcuts, which describe arbitrary points within code
to be manipulated by so called AOP advices. 
As AOP allows to manipulate Java bytecode at runtime without the need to
recompile code, it is perfectly suited to track any code within the server
during processing of requests. 
The server, the used web framework and the developed application do not have
to be modified for this purpose, facilitating high transparency for this
approach. 
From a developer's view, the only difference compared to usual development is
the non-recurring configuration of some additional plugins and the activation of
AOP for the development server as described in section \ref{sec:casestudy}.

When affected web framework parts of a certain point of interest such as UI
processing are analyzed, corresponding AOP pointcuts need to defined. 
When a pointcut is reached during the processing of the web framework, applied
advices will be executed in addition to the original code. 
Although any modifications would be possible here, this approach just aims to
collect required information, generate corresponding metadata and transfer it
to the client. 
For example, if a certain line of a JSF page is parsed and contains a UI
component tag, an AOP advice collects information on the line number, type of
component, set attributes and similar values to generate an adequate
representation of this metadata and include it into the reponse. 
More details on this step are described in section \ref{sec:casestudy} refering
to the JSF sample. 
The definition of pointcuts, advices and adaption of the development environment
has to be done only once, being reusable afterwards by different developers
interested in the same server layer. 

The third question declared above deals with the client-side usage of the meta
data. 
As the approach aims to facilitate the selection of elements within the web page
to obtain information on its sources, the browser has to provide functionality
for the transfered metadata. 
This requires a browser add-on, able to select web page elements and use its
transfered metadata in some way. 
One task is to simply display metadata within the browser plugin in a human
readable format. 
That already affords to recognize affected paths during processing as described
in section \ref{sec:scenario}. 
Furthermore, the browser plugin could perform actions on the affected parts such
as highlighting processed lines of code within the IDE.

\begin{figure}[t]
	\centering
		\includegraphics[width=1.0\textwidth]{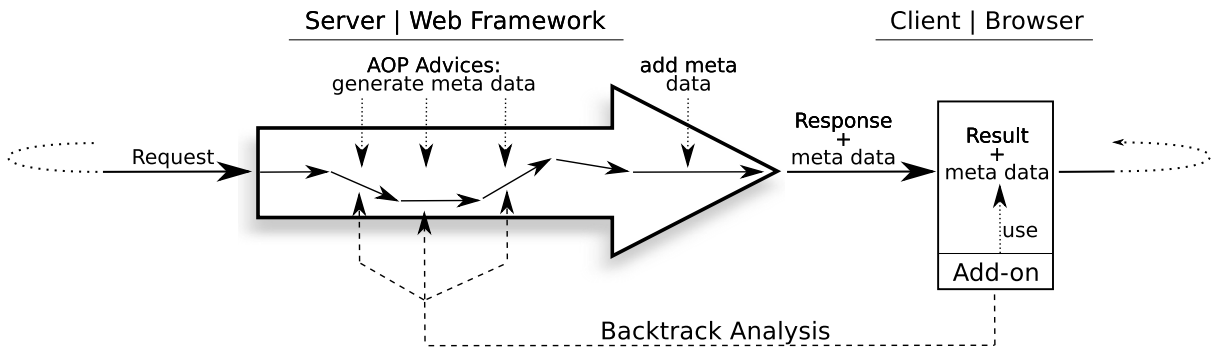}
		\caption{Metadata is generated with AOP, transfered to the client and used by
		a browser add-on to analyze elements. }
	\label{fig:instrumentedGenerationProcess}
\end{figure}
Figure \ref{fig:instrumentedGenerationProcess} shows an overview of
this approach. 
Incoming requests will take a certain path within the web framework as
described in section \ref{sec:scenario}. 
The taken path and arbitrary information as defined by AOP pointcuts will be
observed to generate corresponding metadata. 
That metadata is included into the response and transfered to the client. 
An add-on within the browser makes use of the metadata to display
observed information or even to perform arbitrary actions on the sources as
described in section \ref{sec:casestudy}.

%% file: casestudy.tex
\begin{figure}[t]
	\centering
		\includegraphics[width=1.0\textwidth]{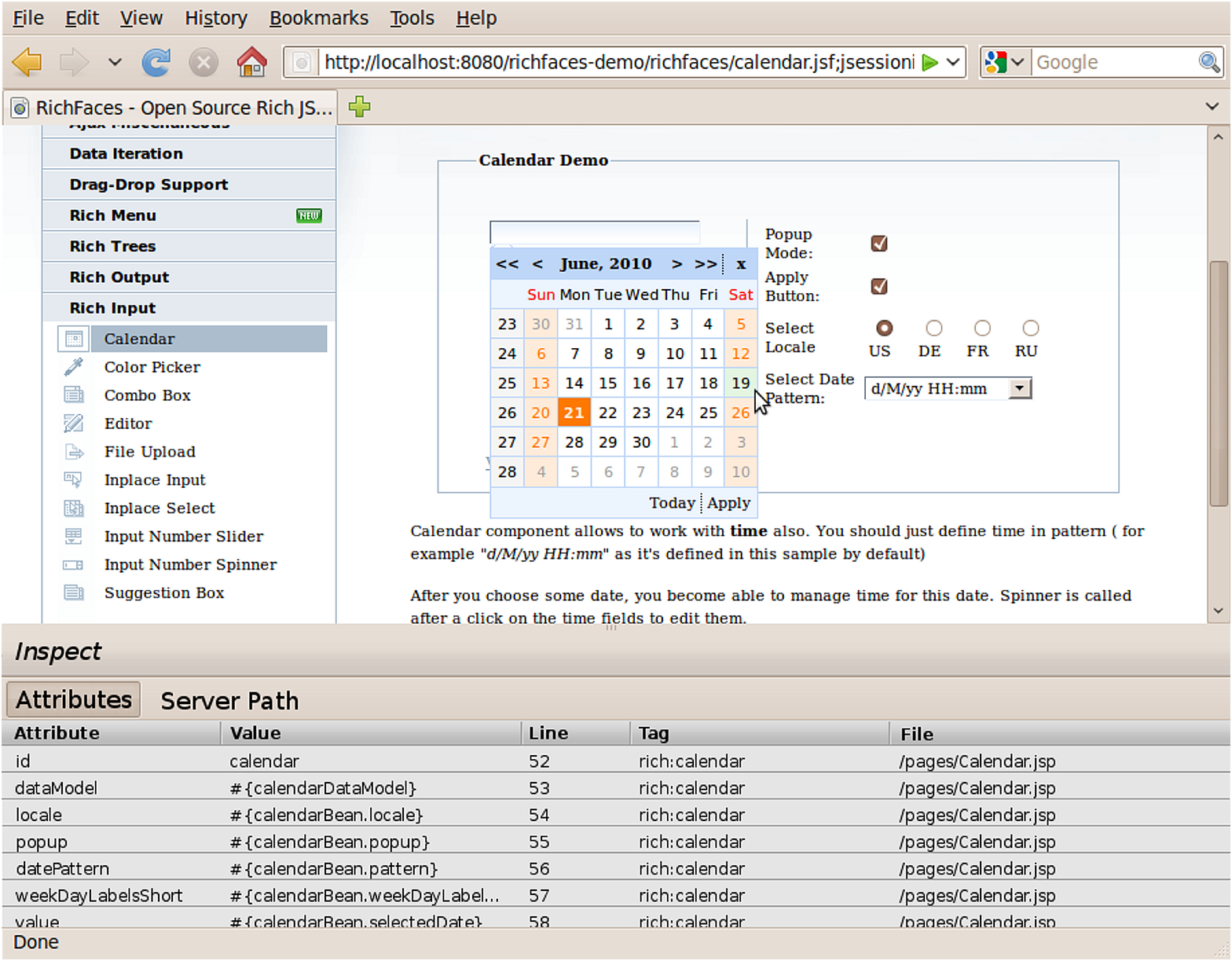}
		\caption{Browser add-on uses metadata on inspected elements to analyze
		server-side processing paths, involved classes, method calls, tags and
		attributes. }
	\label{fig:browserAddOn} 
\end{figure}

Section \ref{sec:scenario} described a scenario with a scope of JSF UI
components and section \ref{sec:approach} dealt with the approach on how to
enable backtracking of server-side information. 
This section describes details of the approach on basis of a 
case study. The case study will generate metadata on a RichFaces demo
application for UI components as shown in figure \ref{fig:browserAddOn}.

To facilitate browser-based analysis for JSF UI components, the points which
shall be observed within the JSF lifecycle have to be identified first. 
That is done considering JSF specification details and by analyzing the source
code of used JSF implementations. 
This case study uses the reference implementation of JSF, Sun JSF RI 1.2 as
well as JBoss RichFaces 3.2.1. 
In addition to that, the case study is based on the JBoss application server 4.2 
\cite{Fleury05JBoss4OfficialGuide}, its contained AOP implementation
\emph{jbossaop} \cite{Fleury05JBoss4OfficialGuide}, the browser Firefox 3 and 
the IDE Eclipse Europa \cite{Daum04ProfessionalEclipse3, Clayberg08EclipsePlugins}. 
An analysis revealed following key points of JSF UI components to be observed:

(1) As described in literature \cite{SchalkBurns06JSFCompleteReference12} and
obvious from source code, JSF UI components consist of at least three parts:
First, a main class represents the UI component and its attributes and is
typically derived from \emph{javax.faces.component.UIComponent} or a subclass. 
Second, a tag class describes how tags and corresponding attributes can be used
within JSF pages and is derived from 
\emph{javax.faces.webapp.UIComponentTag},
\emph{javax.faces.webapp.UIComponentELTag} or subclasses. 
At least, this is true when utilizing JavaServer Pages, the default view handler
technology \cite{SchalkBurns06JSFCompleteReference12} for JSF.  
And finally, a tag library descriptor (.tld) file defines the configuration
of usable tags corresponding to the tag class declared above. 
Since these three parts are mandatory for JSF UI components, they are primary
points to be observed by AOP metadata generators. 

(2) The most important phase referring to the JSF lifecycle depicted in figure
\ref{fig:jsfLifecycle} is the last one called \emph{Render Response}. It parses
requested JSF pages, contained JSF tags are processed by corresponding component
tag and UI classes and renderer classes transform internal component
representations to HyperText Markup Language (HTML). These parts also encompass
line numbers and attributes of the UI components. Observing these parts is
described in detail with code samples below. 

(3) The first JSF lifecycle phase restores UI component states of preceding
requests if any are done to the same page before within a session. 
This step is relevant for observing AJAX communication in general and AJAX 
partial page updates in particular. 

(4) JSF phases 2 to 5 perform some minor tasks referring to UI components such
as validation and conversion of their current values. 
Therefore, they should be observed for advanced information but do not account
for core functionality. 

In a next step, AOP pointcuts must be defined that match the identified points
above to be observed. 
It is advisable to prefer common JSF pointcuts and advices over specific ones
where possible. 
For example, if any JSF component class is derived from the standard JSF class
\emph{UIComponent}, manipulation of this class should be preferred over
component classes specific to RichFaces. 
Thus, an advice can likely be applicable to multiple implementations, e.g. a
different extension than RichFaces. 
However, there might be library specific implementations that do not apply to
intended standard course of action and would not be triggered by common
pointcuts. 
Therefore, extension libraries like RichFaces might require to be observed by
specific AOP pointcuts. 
Considering key point 2 described above, advices and pointcuts can be defined
as shown in listing \ref{listing:pointcut}:
\begin{figure*}
\begin{lstlisting}[label=listing:pointcut,caption={Definition of AOP pointcuts
and advices, which generate metadata during server-side processing of
requests.}] 
<?xml version="1.0" encoding="UTF-8"?>
<!DOCTYPE aop PUBLIC "-//JBoss//DTD JBOSS AOP 1.0//EN" "http://labs.jboss.com/portal/jbossaop/dtd/jboss-aop_1_0.dtd">
<aop>
  <aspect class="de.bkersten.analyze.server.aop.advices.TagAdvice"/>
  <aspect class="de.bkersten.analyze.server.aop.advices.ComponentAdvice"/>
  <aspect class="de.bkersten.analyze.server.aop.advices.RenderAdvice"/>
	
  <bind pointcut="execution(* javax.faces.component.html.*->set*(..))">
    <advice name="setter" aspect="de.bkersten.analyze.server.aop.advices.ComponentAdvice"/> 
  </bind>
  <bind pointcut="execution(* org.richfaces.component.html.*->set*(..))">
    <advice name="setter" aspect="de.bkersten.analyze.server.aop.advices.ComponentAdvice"/> 
  </bind> 
</aop>
\end{lstlisting}
\end{figure*}

Lines 8 to 13 define pointcuts, which describe points within the source code
to be observed according to declared key points. 
These pointcuts observe any execution of setter-methods within classes of UI
components to collect information on their attributes. 
This works because any attribute of an UI component is set by this method
according to the JSF implementation. 
The setter-methods are intercepted by two pointcut definitions, one for
the namespace of standard JSF components and a second for the RichFaces
namespace. 
The pointcuts could be even merged to a namespace pattern like
\emph{*.component.html.*} if there were no conflicts with different libraries.  
Adhering reusability, both pointcuts are bound to the same
advice \emph{ComponentAdvice}, which is triggered when defined setter-methods are called. 
The advice executes additional logic, i.e. it generates meta
data before the intercepted code is continued. 
The \emph{ComponentAdvice} observes creation and restore of components as well
as their attributes to generate corresponding metadata. 
There are other advices for different tasks defined in lines 4 to 6. 
They are bound to different pointcuts not illustrated in listing
\ref{listing:pointcut}.

Listing \ref{listing:advice} shows a code snippet of an advice that is called to
generate metadata on an observed pointcut. 
\begin{lstlisting}[label=listing:advice,caption={AOP advice generating meta
data on observed pointcuts.}]
public Object setAttr( Invocation invocation ) throws Throwable {
	// [...]
	MethodInvocation methodInvocation = (MethodInvocation)invocation;
    
    // [...] collect information
	Object[] arguments = methodInvocation.getArguments(); 
	UIComponent component = (UIComponent)arguments[0];
	int lineNumber = lineNumberObserver.getCurrentLine();
    HttpSession session = (HttpSession)FacesContext.getCurrentInstance().getExternalContext().getSession( true ); 
	// [...] etc.
	
    // [...] generate metadata
	metaData.setUIComponent( component );
	metaData.setId( component.getId() );
	metaData.setLineNumber( lineNumber );
	metaData.setSession( session );
	
	// resume with original source code
	Object result = invocation.invokeNext();
    return result;
}
\end{lstlisting}
Since this advice is called with the execution of setter-methods, an invocation
of the type \emph{MethodInvocation} is passed to the advice to access original
source code and intercepted objects. 
Thus, the UI component currently handled can be read as shown in lines 6 and 7. 
Furthermore, the advice collects any information relevant for this point of
interest as shown for the line number and session in lines 8 and 9. 
Afterwards, collected information is used to generate corresponding metadata as
illustrated in lines 13 to 16. 
Finally, the original course of execution is resumed until the next advice is
called. 
This is done for any pointcut defined for enabled points of interest resulting
in metadata required to be transferred to the client at the end of the process. 
Therefore, the AOP implementation of this case study intercepts rendering
methods of UI components and adds generated metadata as hidden HTML input field
in front of corresponding UI components. 
Optimization for this procedure and specification of metadata processing
details is still subject to future work (cp. section \ref{sec:conclusion}). 

On the client-side, a browser add-on utilizes the 
hidden metadata to display information on inspected elements. 
As depicted in figure \ref{fig:browserAddOn}, a Firefox add-on written in XML
User Interface Language (XUL) and Javascript provides that functionality for
this case study. 
A button \emph{Inspect} is used to select an element and obtain its metadata. 
There are currently two tabs \emph{Attributes} and \emph{Server Path}
available.
According to the samples of listings \ref{listing:pointcut}
and \ref{listing:advice}, the first tab displays the tag and attributes of the
inspected UI component as defined in the JSF page source. 
The second tab \emph{Server Path} displays the path taken within the server
process, i.e. involved classes, methods and line numbers such as
\emph{org.richfaces.component.html.HtmlCalendar},
\emph{org.richfaces.taglib.CalendarTag} or
\emph{org.richfaces.renderkit.html.CalendarRenderer}. 

As described before, this use case is extensible to different points of
interest, AOP aspects and procedures for metadata generation. 
A benefit to notice is that the definition and implementation of AOP pointcuts
have to be done only once and can be reused afterwards. For instance, the
pointcuts developed for this case study could be reused by other developers
interested in analyzing JSF RichFaces UI components. 
Therefore, developers do not need to have intricate knowledge on frameworks or
write a single line of code. 
However, custom implementations have to be created for
different points of interest and web frameworks which have not been handled yet.

%% file: discussion.tex
While this paper focused metadata generation of UI componentes, there are
several different points of interest for observation as already declared in
section \ref{sec:introduction}. 
\begin{figure}[t]
	\centering
		\includegraphics[width=0.9\textwidth]{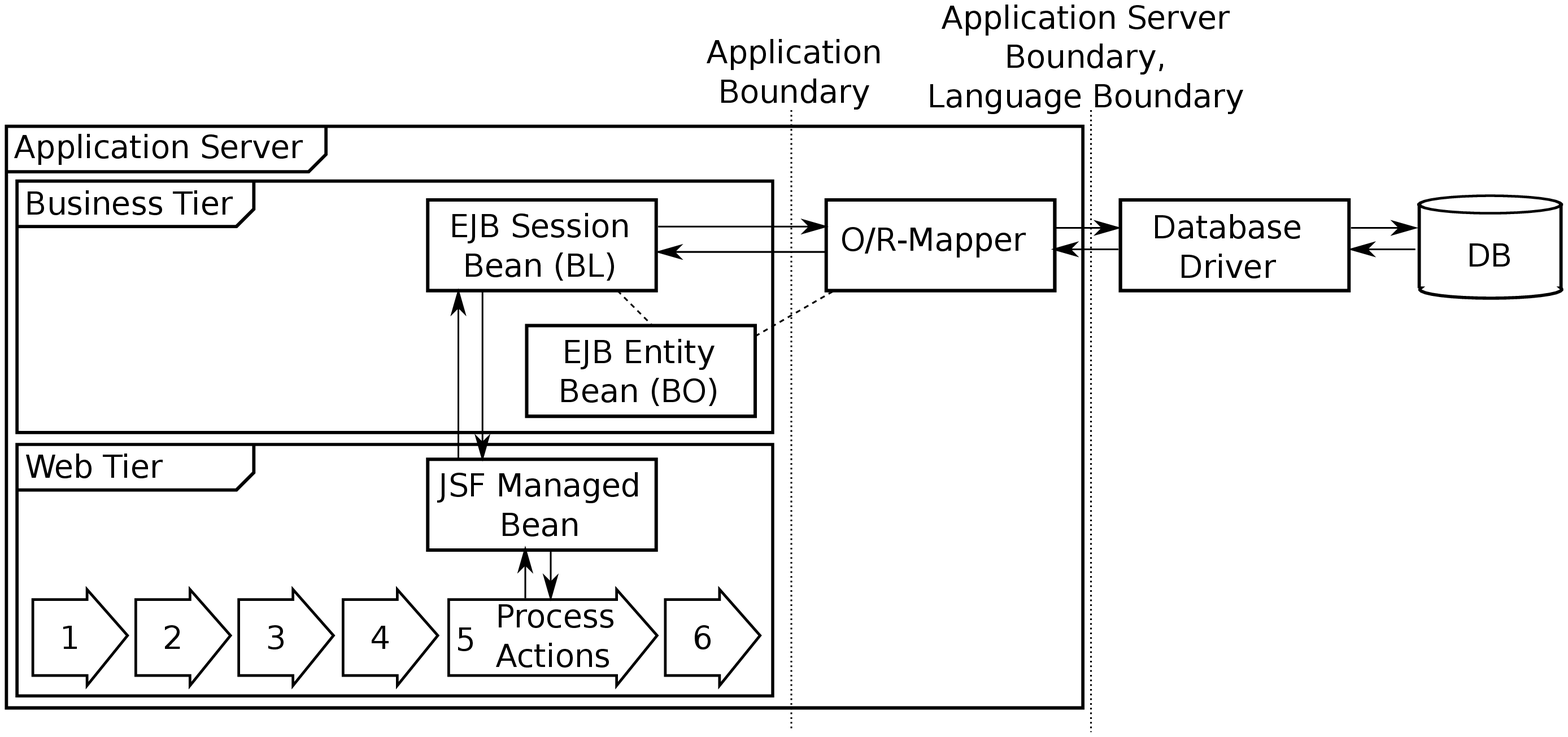}
		\caption{Scenario of JEE business data aggregation to show the applicability
		to different points of interest as well as limitations of the approach. }
	\label{fig:businessData} 
\end{figure}
Figure \ref{fig:businessData} illustrates a sample for observing business data
aggregation. The bottom of the figure shows the six phases of the JSF lifecycle as explained
in section \ref{sec:casestudy}, where phase 5 processes an action to obtain
business data. 
Maintaining the use case of a JEE application, a JSF managed bean
\cite{SchalkBurns06JSFCompleteReference12} within the web application would
delegate the access of business data to an Enterprise Java Bean
(EJB)\cite{EJB3Spec} backend. 
The business logic (BL) is represented by an EJB session bean containing EJB
query language (EJB QL) requests to access a data source. 
This sample application is connected to a database with a preceding
object-relational-mapper (O/R-mapper) and database driver. 
The O/R-mapper transfers business objects (BO, entities) to database tables
(relations) and vice versa. 

On the one hand, this sample shows that the approach of the paper is extensible
and applicable to further points of interest. 
Within the boundaries of the application server, large parts of the
application, web framework and third-party libraries such as the O/R-mapper can
be observed with AOP. 
On the other hand, the sample indicates limitations of the approach, as the
overall system typically relies on external sources such as databases, legacy
systems or native libraries. 
In particular, parts beyond the server boundary or language boundary are not
covered by this approach. 
These parts are difficult to observe, not observable or not
interceptable by AOP aspects. 

In general, AOP expressiveness is a debatable point. 
AOP was chosen since it can easily intercept large parts of the application,
even access third party libraries and provides a transparent approach as
utilized libraries themselves do not have to be changed. 
An optimal solution would use common pointcuts and advices to generate meta
data and would be applicable to multiple implementations once defined. 
Where common pointcuts are not sufficient, more specific pointcuts or advices
could be used inducing some redundancy as explained in section
\ref{sec:casestudy}. 
However, some issues might not be interceptable at all and AOP expressiveness
also depends on the AOP implementation like \emph{AspectJ}
\cite{Laddad08AspectJ} or \emph{jbossaop}. A solution of this problem could be a
manipulation of the used libraries themselves instead of runtime manipulation
with AOP. That would destroy transparency given by AOP and require to use adapted
development libraries different to runtime libraries instead. 
However, with development tool support for library management like Maven
\cite{Sonatype08Maven}, this approach would be acceptable too. 

A complete process for observing and generating metadata as aimed in this
paper should also include observation of client-side code, especially
Javascript. 
In particular, this is true because some web frameworks generate essential
parts on client-side, e.g. the Dojo framework \cite{Russell08Dojo} generating
HTML code for UI components with Javascript. 
Therefore, depending on the used web framework and points of interest this
approach has to be completed with client-side observation. 
However, the paper focused the server-side as there is already work on client
side observation (cp. section \ref{sec:related}), whereas the integration of
the server-side process was not studied yet. 

Finally, the presented approach of metadata generation could be the basis for
different browser-based tools besides analysis. 
For instance, generated metadata on UI components as presented in the case
study of the paper could be used for a visual editor running as browser add-on. 
A browser add-on developed for this case study already facilitates
communication with an IDE, e.g. to manipulate source code. 
Thus, the attributes of UI components could not only be displayed within the
browser but also edited in original source files. 
For the idea of a visual editor, the approach of server-side data within
browser-based tools has the additional benefit to work directly with generated
web pages as finally displayed to the end user. 
It can even handle advanced client-side states such as components only
displayed after execution of Javascript code, e.g. within popup windows. In
comparison, visual editors integrated within IDEs cannot handle these
scenarios but make use of imprecise placeholders instead.

%% file: relatedwork.tex
The first statement of this paper was that browser-based tests are necessary in
addition to different test types. 
There is much work sharing this mindset for different reasons, in particular
with respect to AJAX applications. 
For instance, different test types and techniques, such as white box tests,
black box tests and state-based tests are topic of \cite{Marchetto08AjaxWebTestingTechniques} and
\cite{Marchetto08StateBasedTestingOfAjaxWebApplications} to improve tests of AJAX enabled applications triggering
partial page updates. 
Browser-based tests also have tool support such as provided by Selenium
\cite{Crispin09AgileTesting}. 

Another research topic called \emph{Dynamic Testing} is the automation of these
browser-based tests by generating test cases. 
This is done in \cite{Mesbah09InvariantBasedAutomaticTesting} and
\cite{Raffelt08HybridTestOfWebApplications}, where the first performs checks on
all client-side states and AJAX faults. 
This addresses tracing on client-side, which is not yet included in the approach
of this paper as discussed in section \ref{sec:discussion}. 

All of the approaches mentioned above emphasize test cases to improve detection
of errors. 
However, there is less work for analyzing and locating errors. 
Existing approaches explore analysis in conjunction with modeling
\cite{Halfond08WebModelingForTestAndAnalysis} or deal with different aspects of
complexity for large web applications such as huge site structures and
navigability \cite{Tilley01WebSiteStructure}. 
Browser-based tool support for analysis as provided by Firebug
\cite{Lerner07Firebug} is only supported for generated client-side artifacts
such as HTML, style sheets or Javascript. 

An approach similar to that of this paper is presented in
\cite{Baresi08MonitoringBPELProcesses} which also observes and gathers server-side data during processing with AOP. 
However, the work has a
different focus by monitoring the evolution of BPEL processes of systems at
runtime whereas this paper uses AOP monitoring for analysis of web framework
processes at development time. 

Finally, work related to this paper is research on AOP expressiveness and
quality of pointcuts \cite{Breuel07JoinPointSelectors}. 
That addresses the issue of server parts interceptable by AOP as discussed in
section \ref{sec:discussion}.

%% file: conclusion.tex
In this paper, we presented an extensible approach to generate metadata during
the server-side processing of web requests facilitating browser-based analyses
of web applications at development time. 
Metadata may encompass involved server-side processing units as well as
arbitrary information on different points of interest such as UI component
generation, business data aggregation or application states.  
Both, observation of points of interest as well as generation of metadata is
done with AOP to provide a transparent approach easily covering large parts of
server-side processing. 
On client-side, a browser add-on can be used to inspect HTML elements within
generated web pages to obtain information collected on the server-side. 
Thus, a developer can display useful information and comprehend server-side
processing of defective elements within generated pages to enhance error
locating. 
Instead of debugging issues manually, lots of key points can already be
displayed within the browser. 

The paper presented a case study, serving as feasibility study and for more
detailed explanations. 
It used the Java web framework JSF and focused metadata generation of JSF
UI components. 
The case study demonstrated that the approach is working generally and
applicable to large parts of server-side processing. 
Nevertheless, there are open questions considering different frameworks,
non-standard processings and different points of interest which are still
subject to future work. 

In addition, future work also encompasses a specification of required common
attributes and process details for metadata generation to facilitate
interfaces for additional implementations of different libraries and points of interest.